\documentclass{acm_proc_article-sp}

\usepackage{algpseudocode}
\usepackage{amsmath}
\usepackage{bbm}
\usepackage{epsfig}
\usepackage{caption}
\usepackage{subcaption}
\usepackage{pdfpages}
\usepackage{graphicx}

\usepackage{array}
\newcolumntype{P}[1]{>{\centering\arraybackslash}p{#1}}
\newcolumntype{M}[1]{>{\centering\arraybackslash}m{#1}}

\newcommand{\pro}{\emph{Control}}
\newcommand{\anti}{\emph{Rights}}

\usepackage{todonotes}
\newcommand{\Note}[2]{}
\def\sharedaffiliation{%
\end{tabular}
\begin{tabular}{c}}

\newcommand{\removed}[1]{}

\begin{document}

\title{After Sandy Hook Elementary:\\A Year in the Gun Control Debate on Twitter}

\numberofauthors{5}
    \author{
\alignauthor
Adrian Benton$^1$\\
       \email{adrian@cs.jhu.edu}
\alignauthor
Braden Hancock$^2$\\
       \email{braden.hancock@stanford.edu}
\alignauthor
Glen Coppersmith$^3$\\
       \email{glen@qntfy.com}
\and
\alignauthor
John W. Ayers$^4$\\
       \email{ayers.john.w@gmail.com}
\alignauthor
Mark Dredze$^{1}$\\
        \email{mdredze@cs.jhu.edu}
      \sharedaffiliation
       \affaddr{$^1$Center for Language and Speech Processing, Johns Hopkins University, Baltimore, MD} \\
   \affaddr{$^2$Department of Computer Science, Stanford University, Stanford, CA}\\
   \affaddr{$^3$Qntfy, Crownsville, MD}\\
   \affaddr{$^4$Graduate School of Public Health, San Diego State University, San Diego, CA}\\
           }

\maketitle
\begin{abstract}

The mass shooting at Sandy Hook elementary school on December 14, 2012 
catalyzed a year of active debate and legislation on gun control in the United States.
Social media hosted an active public discussion where people expressed
their support and opposition to a variety of issues surrounding gun legislation.
In this paper, we show how a content-based analysis of Twitter data can 
provide insights and understanding into this debate. 
We estimate the relative support and opposition to gun control measures, along with
a topic analysis of each camp by analyzing over 70 million gun-related tweets from 2013.
We focus on spikes in conversation surrounding major events related to guns throughout the
year.
Our general approach can be applied to other important public health and political issues 
to analyze the prevalence and nature of public opinion.

\end{abstract}

\section{Introduction}

Gun control in the United States is a major public policy issue that has
polarized US society \cite{schildkraut2014}.
Although public opinion has been strongly in favor of stricter gun control policies
for over two decades \cite{vernick1993}, federal gun control legislation has been
a hotly contested issue meeting little legislative success, where even local restrictions
have been met with opposition \cite{vizzard2015}. Insofar as public opinion affects
the bills debated and passed into law, accurately gauging public opinion and
salience on the various issues associated with gun control is important to
inform the legislative process \cite{burstein2003,monroe1998,page1983}.

Public opinion is typically estimated through written or telephone surveys
where subjects are asked to share their level of approval of different
policies up for debate \cite{barry2013,kleck2009}.  Assuming the population
is uniformly sampled and that subjects are able and willing to divulge their
true beliefs, these are reliable proxies for public opinion.  Gun control polls are
often conducted over the phone and ask respondents about their gun ownership, as well as
opinions on different forms of gun control legislation (e.g., ``Saturday night special'' bans,
assault weapon bans, national firearm registration, universal background checks) \cite{vernick1993}.

However, traditional surveys have a number of drawbacks, including limitations on the response types
and cost restrictions on producing timely results. These limitations are well known in the public
health realm where surveys, a critical data source for a variety of public health topics,
are facing increasing feasibility challenges. As a result, researchers have turned to new data sources,
such as search queries\footnote{\url{https://www.google.com/trends/story/US_cu_ZM8QflEBAAAlMM_en}} \cite{ginsberg2009detecting} and social media \cite{Dredze:2016db}. 
Social media has been used to estimate public opinion on a range of topics, including 
political sentiment \cite{bermingham2011,oconnor2010,sang2012,tumasjan2010}
and a range of public health topics \cite{Benton:2016dn}, including gun control \cite{Ayers:2016uo}.
Some work has looked at gun control tweets, but has focused on argument framing and not
measuring public opinion \cite{stefanone2015}. 

Issues of gun control came to the forefront of national discussion with
the mass shooting at Sandy Hook Elementary School in Newtown, Connecticut on December
14, 2012.  This tragedy followed six months after
another mass shooting in an Aurora, Colorado movie theater, and prompted a concerted
effort to pass stronger gun restrictions at the federal level.  In April, 2013, a
bill to expand background checks was defeated in the senate, ending federal legislative efforts. Failure to pass national
gun control legislation led many states, including Colorado and Connecticut, to pass their own gun
control bills.

Public opinion played a major role throughout this time period, where discussions of gun
control on social media rose in prevalence and prominence. The richness of social media data,
where we have both overall prevalence, content and location data, presents new opportunities
for analyzing and understanding the nature of public opinion surrounding guns. 

We present an analysis of gun-related Twitter data from all of 2013, over 70 million tweets
in total. 
We focus on two main questions: 1) Do Twitter conversations in support of or opposition to 
gun control reflect public opinion as measured by traditional surveys?
2) What events generate online activity
from gun control supporters and opponents and how do the arguments and issues discussed change
in response to these events?
While there has been significant work addressing our first question in regards to other 
topics of public opinion \cite{oconnor2010,Benton:2016dn}, the second question 
gives us a new framing in terms of
social media studies; we are concerned with \textbf{what}
social media users are saying about gun control, in addition to how many people are saying it.

\section{Methods}

Our data set contains 70,514,588 publicly-available tweets collected using the Twitter
streaming API based on keywords and phrases associated with guns or gun
control in the United States: {\tt gun, guns, second
amendment, 2nd amendment, firearm, firearms}.
Our collection covers just over one year, starting on December 16, 2012
(two days after the Sandy Hook shooting) and ending on December 31, 2013.

\begin{table}[t]
\centering
\small
\begin{tabular}{| M{2.4cm} | M{5.25cm} |} \hline
  \textbf{Keyword type} & \textbf{Keywords} \\ \hline
  General         & gun, guns, second amendment,
                              2nd amendment, firearm, firearms \\ \hline
  \pro{}  & \#gunsense, \#gunsensepatriot, \#votegunsense,
                              \#guncontrolnow, \#momsdemandaction,
                              \#momsdemand, \#demandaplan, \#nowaynra,
                              \#gunskillpeople, \#gunviolence,
                              \#endgunviolence \\ \hline
  \anti{} & \#gunrights, \#protect2a, \#molonlabe,
                              \#molonlab, \#noguncontrol, \#progun,
                              \#nogunregistry, \#votegunrights,
                              \#firearmrights, \#gungrab,
                              \#gunfriendly \\ \hline
\end{tabular}
\caption{Keywords used to collect tweets are listed as \emph{General}
         keywords, and hashtags suggesting a \pro{} or
         \anti{} gun control stance.}
         \label{table:kws}
\end{table}

We identified hashtags indicative of support for (\pro{}) or opposition
to (\anti{}) gun-control as a rough estimate of sentiment towards gun control.
These hashtags were strongly associated with either the \pro{} or \anti{} gun
control positions. We obtained this list by examining the most
popular hashtags in a subset of our data and selecting those that
strongly indicated either one of these positions.
Table \ref{table:kws} shows these hashtags: 11 for
\pro{} and 11 for \anti{}. A tweet was labelled as \pro{}
gun control if it contained more \pro{} hashtags than \anti{},
and vice-versa for \anti{} tweets.  A total of 304,142
tweets were labelled as \pro{} and 125,936 as \anti{} using this
method.  Although only about 0.6\%
of tweets were coded with gun control stance, this labelling method resulted
in a high precision coding of tweets by gun control stance.
We leave to future work statistical methods that identify gun control sentiment
of a larger percentage of our data
\cite{jiang2011target,vijayaraghavan2016deepstance}.

Our content analysis of the sentiment coded tweets relies on latent Dirichlet
allocation (LDA) \cite{blei2003},
a data-driven probabilistic topic model that can identify the major thematic elements in a text corpus. 
Topic models infer the parameters of a probability distribution with
Bayesian priors, producing for each topic a distribution 
over the words in the corpus. Reviewing the most probable words for each topic is a common 
technique for establishing a semantically grounded label for the topic.
Additionally, the model assigns a distribution over topics to each document (tweet),
which enables the tracking of topic proportions in a corpus over time \cite{griffiths2004}.
Topic models have become popular tools for analyzing text data in social science \cite{grimmer2010},
the humanities \cite{newman2006,mimno2012} and health \cite{paul2011, paul2012}, with numerous examples of applications to Twitter data \cite{hong2010,ramage2010,zhao2011}.

We sub-sampled 6 million tweets (8.5\% of the total collection) to train an LDA model, and then
used the learned parameters to infer document specific topic distributions for each tweet.
Tweets were tokenized by non-alphanumeric characters into unigrams and filtered using a 
stopword list specific to Twitter. We retained the 40,000 most frequent word types for learning.
We used the LDA implementation in Mallet \cite{mccallum2002} and tuned model parameters on a held out
set of 1 million tweets to maximize model log-likelihood. We swept the number of topics
from 25 to 500, and the document-topic Dirichlet prior hyper-parameter $\alpha$ from 0.25 to 10 (with an asymmetric prior.)
We used Mallet's parallel Gibbs sampler with a burn-in of 100
iterations, 500 total iterations, with hyper-parameter optimization
every 10 iterations. Our tuned model used 
an initial $\alpha=1$ and 250 topics.  The final model was then used to
infer topics for the entire corpus using 200 sampling iterations.

We obtained a location for each tweet using Carmen \cite{dredze2013a}, a
high-precision geocoder for Twitter based on a user's profile. Wherever possible, we obtained
the US state associated with a tweet. We chose to rely on an automatic geocoder since the proportion of tweets
with location information provided by Twitter was small (around 1-2\%.)

Using the sentiment coded tweets and their inferred topic distributions, we measured the following trends.
1) The overall number of gun-related tweets
for each day and week during 2013. 2) The number of
\pro{} and \anti{} messages for each day and week. Since the
overall Twitter volume remained relatively stable in 2013, our counts are not
normalized. 3) The most likely topics associated with \pro{} or \anti{} tweets over the entire corpus, as well as for each week.
This gives us a fine-grained look at which topics were
discussed by each gun control camp for each week.
We compute these trends for both the entire United States and for each US state.

\section{Results}

\subsection{Comparison with Polling Data}

\begin{table}[t]
\centering
\small
\begin{tabular}{| l | l | l | l |} \hline
  Alaska & 4/26/2013 &   Montana & 6/23/2013 \\ 
  Arizona & 4/26/2013 &   Nevada & 4/26/2013 \\ 
  Arkansas & 5/23/2013 &   North Carolina & 5/1/2013 \\ 
  Georgia & 5/23/2013 &   North Carolina & 7/14/2013 \\
  Georgia & 8/5/2013 &   Ohio & 4/26/2013 \\ 
  Iowa & 6/7/2013 &   Ohio & 8/19/2013 \\ 
  Louisiana & 5/1/2013 &   Tennessee & 5/23/2013 \\ 
  Louisiana & 8/19/2013 &   Texas & 7/1/2013 \\ 
  Michigan & 6/2/2013 &   Virginia & 7/14/2013 \\ 
  Minnesota & 5/19/2013 &   Wyoming & 7/21/2013 \\ 
  \hline
\end{tabular}
\caption{Description of the states that were polled by Public Policy Polling,
         and the date they were polled.  Dates are the last
         day the poll was conducted.}
         \label{table:polls}
\end{table}

\begin{figure}[t]
  \centering
  \includegraphics[width=2.5in]{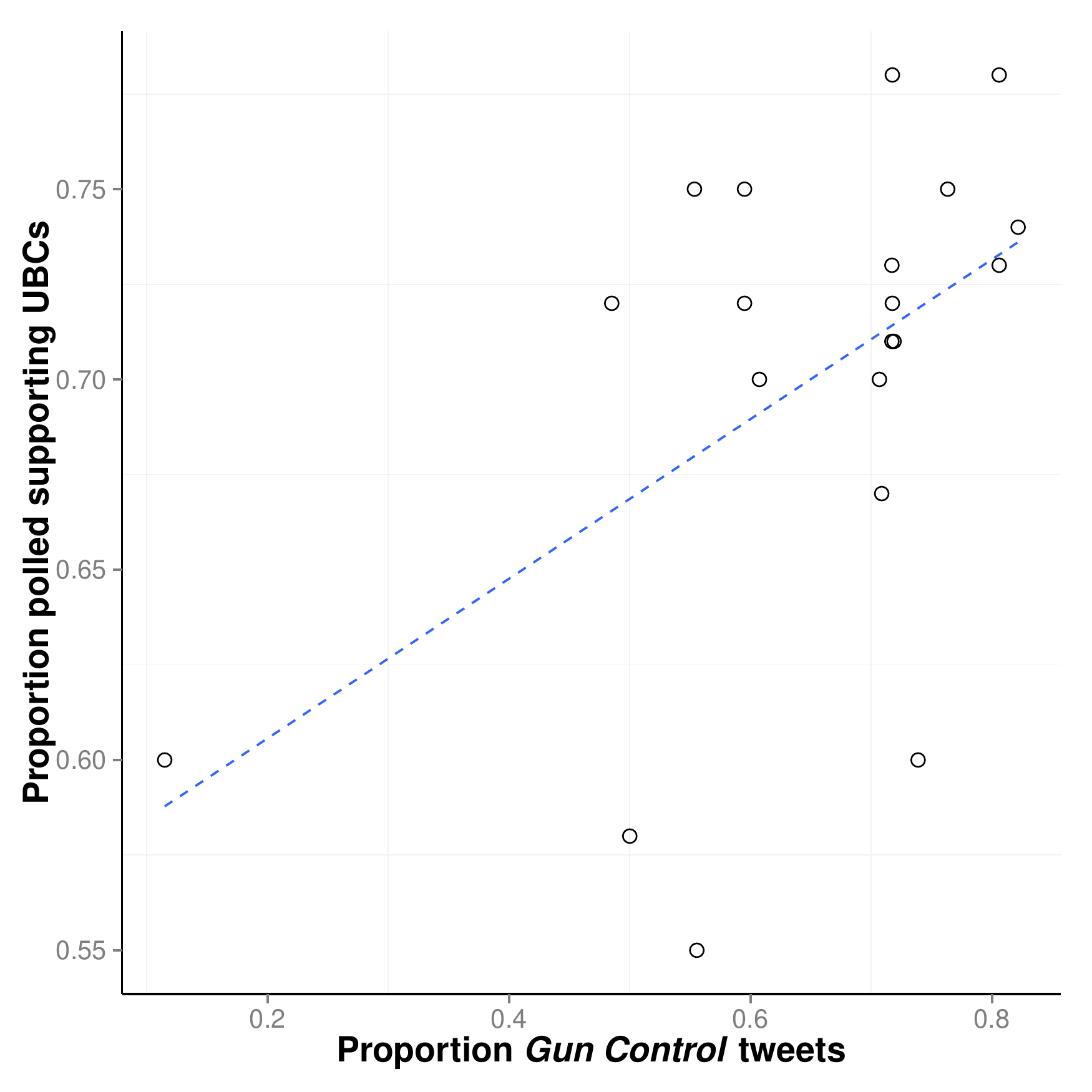}
  \caption{Proportion of \pro{} gun control tweets,
           over all \pro{}/\anti{} tweets from that state, against the
           percent polled in that state supporting universal background
           checks.
          }
  \label{figure:statePollsToTwitter}
\end{figure}

We begin by measuring the ability of Twitter to track gun related opinions as compared
to results from traditional survey methods.
We obtained US state level polling for 16 states
gathered between April 4, 2013 and August 19, 2013 by Public Policy
Polling\footnote{This polling firm ({\tiny \url{http://www.publicpolicypolling.com}}) has a pollster rating of B+ according
to FiveThirtyEight's Pollster Ratings ({\tiny \url{http://projects.fivethirtyeight.com/pollster-ratings/}}), although this rating
is based on US election polls.} -- a total of 20
polls.
The state and date of each poll is included in Table \ref{table:polls}.
Our sentiment coding technique identified 304,142
tweets as \pro{} and 125,936 \anti{}. Of these tweets, a total of 165,360
(38\%) were geocoded with a US state.

While our sentiment coding of tweets was for a coarse \pro{}/\anti{} position on gun control, the polls
do not directly ask this question. Therefore, as a proxy we selected the following question which appeared in all polls:
\emph{``Would you support or oppose requiring
background checks for all gun sales, including gun shows and the Internet?''}.

We used the proportion of ``yes'' answers from each poll as the value for
each US state. For states that had two polls, we used each poll as a separate
data point in our correlation.
For Twitter, we measured the proportion of \pro{} tweets over the
number of both \pro{} and \anti{} tweets for each US
state over our entire collection.  Due to data sparsity, we did not limit the tweets to consider only
those from the time period the poll was taken.

We obtained a Pearson correlation coefficient of 0.51 between our two variables: -- proportion
``yes'' in state polls and proportion of \pro{} tweets.
Figure \ref{figure:statePollsToTwitter} displays the least-squares fit
between these two variables, with an $R^2$ value of 0.22. This is a
reasonably strong relationship between the variables, demonstrating that
relative proportion of buzz in gun conversations on Twitter are
reflective of opinions of the actual population.

This reasonably strong relationship was obtained even with several important
limitations on our method.
First, public opinion varied over time \cite{gallup2001}, yet these state polls capture just a single point in time,
and different time periods at that. The time period of our Twitter data was mismatched to these polls,
in that we used tweets from the entire corpus instead of restricting them to the time when the poll was
conducted. Doing so would have yielded too few tweets, though future work that expanded our 
sentiment classifier method could address this problem.
Second, we were only able to obtain polls and sufficient tweets for some US
states, which reduces our ability to validate this method over the entire
United States. Even though this was a major issue in US politics for a sustained period of time,
polls were not conducted for every state. Third, gun control opinions can be complex, yet we are
measuring only a coarse level of sentiment. The complex opinions expressed on Twitter
may not map directly to our selected question. Fourth, we counted tweets, {\em not} 
the number of accounts tweeting. A single prolific account could bias our estimates. 
Finally, additional errors could be
introduced by the accuracy of the geocoder, Twitter's representativeness of
the US population, and the biases and sampling errors inherent in surveys.
Despite these limitations, the obtained correlation is a strong indicator
of the value of Twitter data for opinion analysis.

\subsection{Opinions Surrounding Events}

\begin{figure}[t]
  \centering
  \includegraphics[width=2.5in]{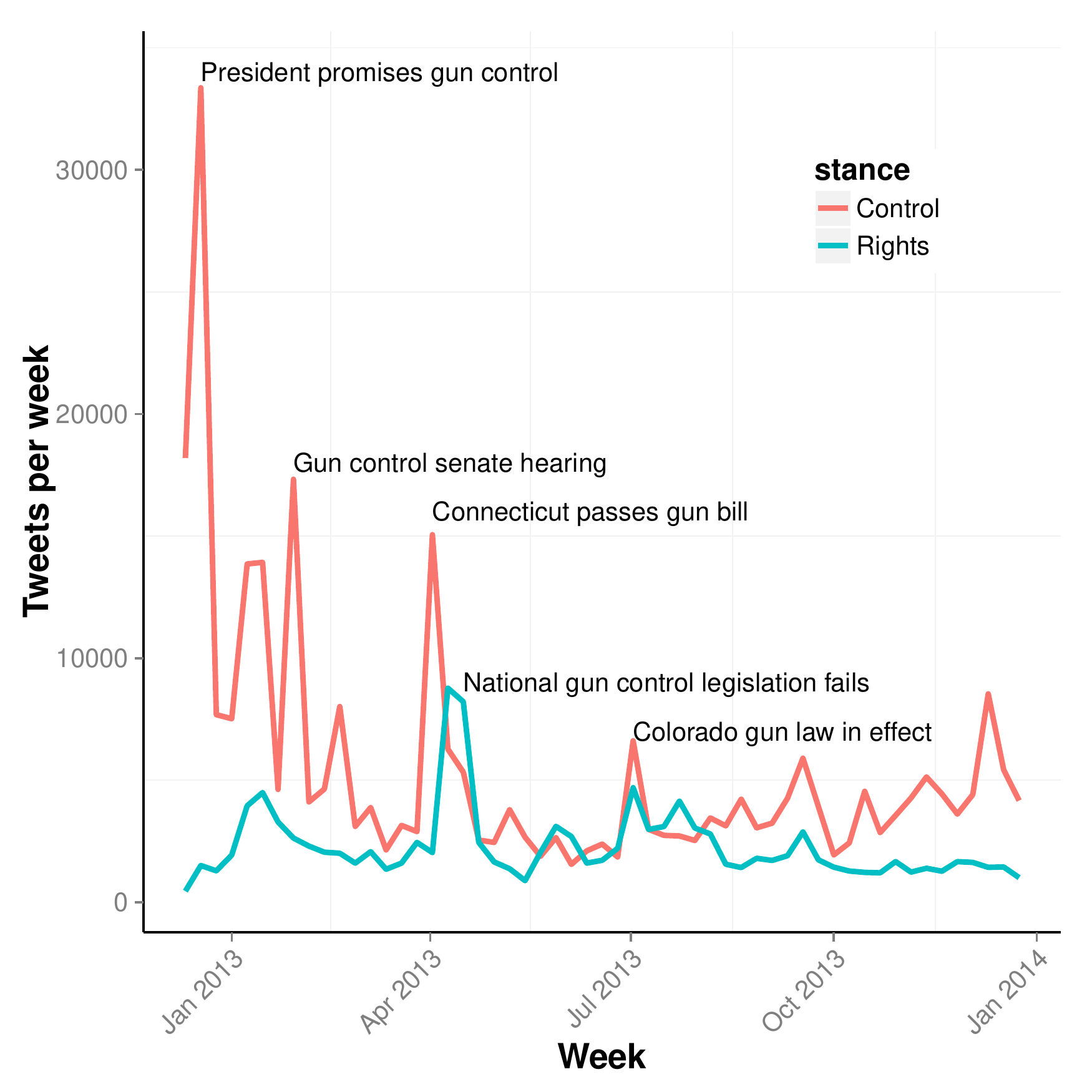}
  \caption{Number of \pro{} and \anti{} tweets over time.  Events
           of interest are annotated above spikes in activity.}
  \label{figure:proAntiTweetsPerWeek}
\end{figure}

We next contextualize opinions as expressed on Twitter within the context of major gun control related events during 2013.
We identified significant spikes in activity using the weekly aggregated statistics of \pro{} and \anti{} tweets.
For each spike, we used a historical news collection to identify major gun
related events corresponding to the spike. Figure \ref{figure:proAntiTweetsPerWeek} displays Twitter 
traffic per week by \pro{} and \anti{} with large spikes annotated with co-occurring events.  Events
of note are:
\begin{itemize}
  \item President Obama promises stronger national gun control
        legislation -- \pro{} tweets spike  (December
        19, 2012)\footnote{\tiny \url{http://nyti.ms/1GfO4jo}}.
  \item The first gun control senate hearing featuring appearances from
        Gabrielle Giffords and Wayne LaPierre -- \pro{} tweets spike (January 30,
        2013)\footnote{\tiny \url{http://www.huffingtonpost.com/2013/01/30/gun-control-hearing_n_2580691.html}}.
  \item Connecticut passes strict gun control legislation in response to
        the Sandy Hook shooting -- \pro{} tweets spike (April 4,
        2013)\footnote{\tiny \url{http://www.huffingtonpost.com/2013/04/04/connecticut-gun-control-sandy-hook-law_n_3011625.html}}.
  \item A compromise is reached over the gun control
        bill, significantly weakening the 
        bill (April 10, 2013) \footnote{\tiny \url{http://nydn.us/1G3P3p4}}.  Subsequently the push for stricter
        gun control was defeated in the senate (April 17, 2013) \footnote{\tiny \url{http://nyti.ms/185ffzu}}.
        \anti{} activity spikes in both cases.
  \item Colorado gun control legislation banning high-capacity magazines goes
        into effect (July 1, 2013) \footnote{\tiny \url{http://www.huffingtonpost.com/2013/07/01/gun-control-colorado_n_3528397.html}}.  Tweets
        increase for
        both \pro{} and \anti{} advocates, although to a lesser
        degree than when national gun control legislation was being
        debated.
\end{itemize}

In our corpus, on average, \pro{} tweets are much more common. \anti{} tweets eclipsed those of \pro{}
when gun control legislation failed to pass in April.

\subsection{Major Topics of Discussion}
\begin{table}[t]
\tiny
\centering
\begin{tabular}{| M{0.4cm} | M{0.5cm} | M{4.5cm} | M{1.5cm} | } \hline
  \textbf{\#} & \textbf{$prob$} &
    \textbf{Representative tokens} & \textbf{Label} \\ \hline
  237 & 0.222 & violence action sense common demand moms muses laws sign
                momsdem and vote congress momsdemandaction gt retweet gunsense
                house republicans prima & ``Common sense'' gun laws
                \\ \hline
  136 & 0.129 & nra amp safety owners people control laws lobby gop violence
                manufacturers responsible don industry americans children
                support money fear congress & NRA
                \\ \hline
  7   & 0.108 & violence barackobama president reduce end plan obama
                demandaplan congress america nowisthetime time support amp
                newtown protect kids action demand agree & National gun legislation
                \\ \hline
  212 & 0.077 & nj pjnet anow tcot momsdemand nra amendment million amp
                rights liberty gunsense firearms owners teeth criminals
                abiding constitution control people & Mix of hashtags
                \\ \hline
  57  & 0.051 & americans died violence wars child deaths fact combined
                america killed barackobama home die times death newtown
                iraq people amp accidental & Domestic violence $>$ foreign violence
                \\ \hline
  246 & 0.040 & background checks check buy sales senate people amp universal
                show nra don private ill shows buying criminals senators pass
                loophole & Universal background checks
                \\ \hline
  120 & 0.039 & crime laws rate control murder states violent deaths violence
                ownership study rates related country highest piersmorgan
                homicides uk murders australia & Model gun control policy
 
                \\ \hline
  48  & 0.033 & killed deaths year children people americans newtown america
                violence million firearms amp related daily firearm women die
                american murders suicide & Domestic violence
 
                \\ \hline
  211 & 0.020 & nra demandaction control mcdonalds newtown paulstewartii
                breakfast whatwillittake bring rank sells respond free support
                challenged pete virginia high mcauliffe polls & Boycott
                 \\ \hline
  216 & 0.016 & make people america safer don safe feel free country live
                world healthcare children piersmorgan nra control society
                protect kill kids & Safety
                 \\ \hline
\end{tabular}
\caption{Top 10 topics ranked by $Prob(topic | \pro{})$.}
\label{table:probProTopic}
\end{table}

\begin{table}[t]
\tiny
\centering
\begin{tabular}{| M{0.4cm} | M{0.5cm} | M{4.5cm} | M{1.5cm} | } \hline
  \textbf{\#} & \textbf{$prob$} &
    \textbf{Representative tokens} & \textbf{Label} \\ \hline
  69  & 0.258 & tcot nra ndamendment tlot guncontrol tgdn control gunrights
                obama amendment registry pjnet protect national sentedcruz
                stand teaparty agree support nogunregistry & Conservative hashtags/gun registry
                \\ \hline
  121 & 0.164 & state law texas laws firearms control nra carolina rated
                connecticut york afriendly friendly north gov bill colorado
                leave rick move & State gun laws
                                \\ \hline
  170 & 0.163 & tcot nra tgdn teaparty tlot pjnet ccot guncontrol lnyhbt gop
                ocra control amendment rkba gt sot obama bcot atomiktiger
                freedom & Conservative hashtags, misc.
                \\ \hline
  6   & 0.084 & latest man robber store news armed home police suspect robbery
                woman guncontrol clerk pulls shoots nra bank owner homeowner
                eqlf & Gun anecdotes (defense)
                \\ \hline
  5   & 0.030 & amendment rights nra amp constitution party support obama
                protect st tea america defend owners people tcot don freedom
                defending protecting & Second amendment
                                \\ \hline
  227 & 0.024 & control bill senate sen filibuster vote feinstein reid senator
                amendment senators paul legislation cruz gop voted harry rand
                dianne ted & Cruz, Paul, \& Reid filibuster
                \\ \hline
  75  & 0.018 & ban weapons assault treaty bill senate arms amendment control
                nra trade obama democrats sign feinstein amp firearms owners
                national registry & Assault weapons ban
                \\ \hline
  225 & 0.014 & show amp day pm today range talk club tonight shooting tomorrow
                firearms night music weekend load saturday live free & Entertainment
                \\ \hline
  2   & 0.013 & control bloomberg mayor nra obama group anti laws michael
                mayors ad illegal push nyc york campaign sheriffs pro million
                state & Bloomberg pro gun control ads
                \\ \hline
  99  & 0.011 & control service secret matter kid toy shoots doesn obama die
                pretend men nra clint agent policy strict batman reagan comics
                & Gun control opinion (bongino)
                \\ \hline
\end{tabular}
\caption{Top 10 topics ranked by $Prob(topic | \anti{})$.}
\label{table:probAntiTopic}
\end{table}

\begin{figure*}[t]
  \centering
  \includegraphics[width=0.48\linewidth]{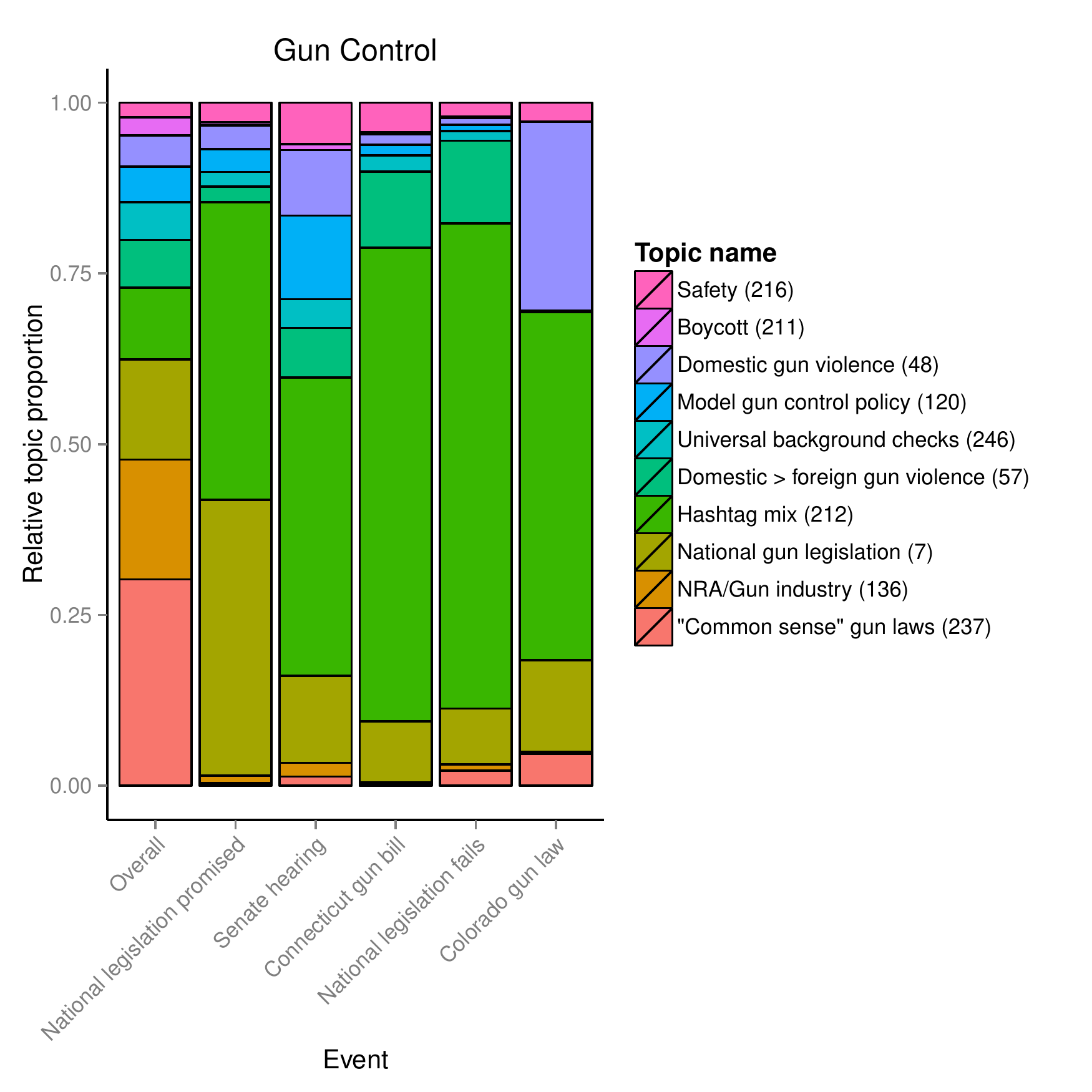}
  \includegraphics[width=0.48\linewidth]{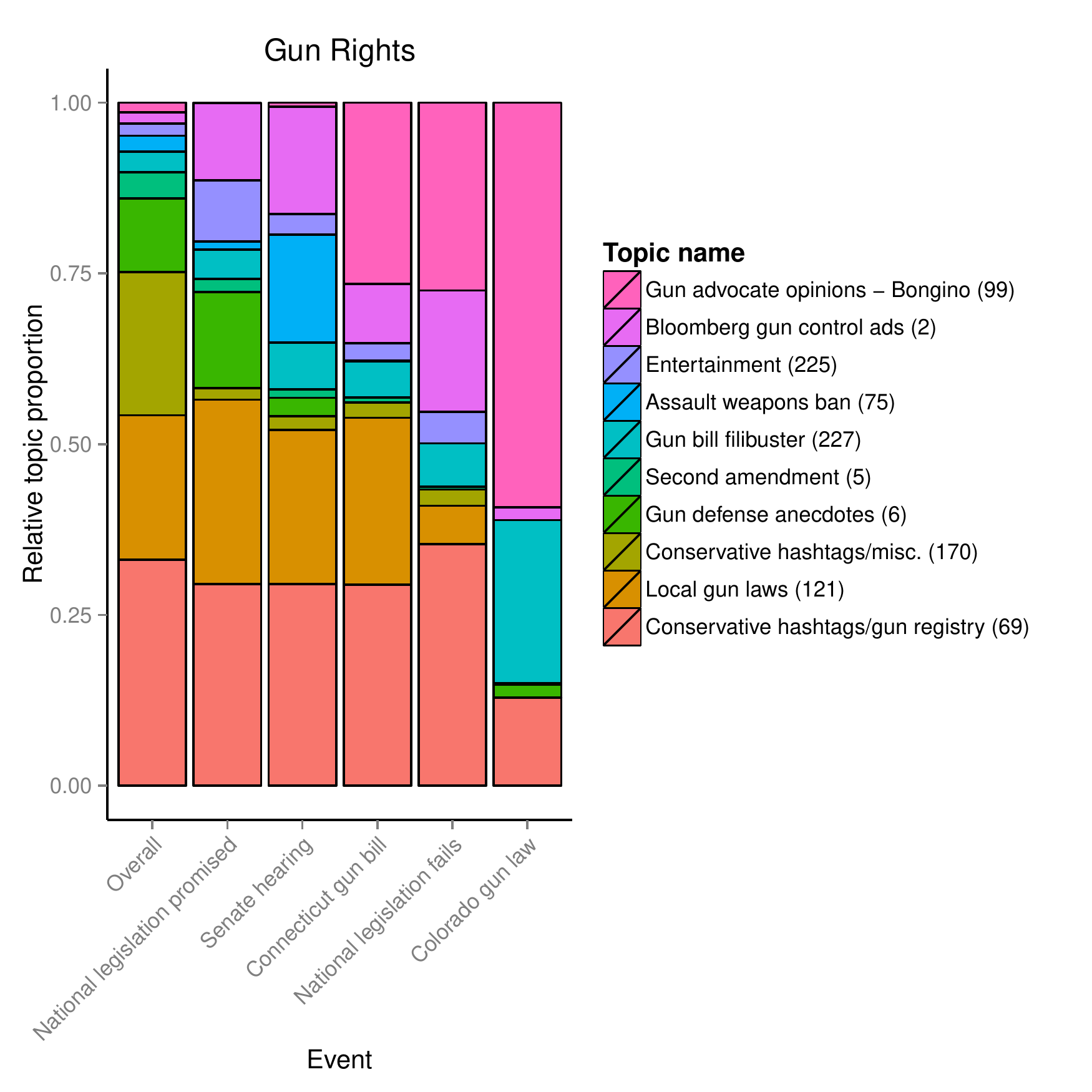}
  \caption{Relative proportion of top 10 \pro{} (left) and
           \anti{} (right) topics overall and during specific events.  The topic number
           is indicated in the legend in parentheses.}
  \label{figure:proAntiTopicsPerEvent}
\end{figure*}

Beyond detecting the overall sentiment surrounding each event, we
characterized the content of each side by examining the topics discovered by
the topic model.
We selected the ten most likely topics for both the \pro{} and
\anti{} tweets over the entire time period: 
304,142 \pro{} tweets (2,802,636 tokens)
and 125,936 \anti{} tweets (1,265,765 tokens). Tables
\ref{table:probProTopic} and \ref{table:probAntiTopic} show these topics, 
their likelihood,
the most likely words, and our assigned label.

These topics and their relative order summarize the
main thrusts of the conversation for both the \pro{} and \anti{}
group.
For example, topic 237 centers
around the group ``Moms Demand Action for Gun Sense in America'' and topic
246 around universal background checks -- both topics prevalent in
\pro{} tweets.  Topic 6 discusses armed robbery
(presumably as an argument against new gun restrictions that would prevent
citizens from protecting themselves) and topic 5 contains language
indicative of political conservatives and second amendment rights
advocates, in general.  

We next contextualized these topics within the events described above.
We computed the distribution over the 10 topics for the
\pro{} tweets
and the 10 topics for the \anti{} tweets in the week around the event. 
By comparing how the usage of these topics change for each event,
we can compute the dominant topics of conversation around each event.
Figure \ref{figure:proAntiTopicsPerEvent} shows
the \emph{relative} proportion of the top 10 topics, \emph{overall} for the
set of \pro{} and \anti{} tweets independently, as well as their
proportion
during each of the events. Topics are ordered by the their relative proportion over
all \pro{} or \anti{} tweets.

When President Obama initially promised federal gun control legislation,
gun control advocates tweeted much more frequently about it, but this was not
as prevalent during most other events, or overall.  Universal background
checks and models of more restrictive gun control policy are also mentioned
much more frequently during the first senate hearing on gun control.

When federal gun control legislation was first promised, gun rights 
tweets centered mostly around self-defense applications and state
laws permitting carrying guns.  During the first senate hearing on gun
control, discussion also focused more on restrictions on assault weapons.
As time progressed, former secret service agent and Republican political
candidate, Dan Bongino became more vocal about gun rights.  This is
reflected in a greater proportion of tweets mentioning him.

\section{Discussion}

By analyzing a year's worth of tweets on guns in the
United States, we find variation in each side's reaction to
gun related events, as well as variation in the arguments cited by
each group during events of interest.  \pro{} advocates are very vocal early
on in the debate when national legislation is still a possibility, but
die down later on.  From Figure \ref{figure:proAntiTopicsPerEvent}, it
is clear that a large proportion of this chatter was about national gun
control legislation (Topic 7).  \anti{} advocates became more vocal once
the national legislation for universal background checks failed in
congress, and much of their subsequent discourse focused on an assault
weapons ban (Topic 75), the senate filibuster (Topic 227),
and political candidate and gun rights advocate Dan Bongino (Topic 99).

We believe that this style of social media analysis is a complement to
traditional polling techniques, which typically gauge opinion on a
small set of issues.  By fitting a topic model to the entire collection of
gun-related tweets in 2013, we are able to identify salient issues and
arguments for both camps, which researchers may not have identified as
relevant, a priori.  Most importantly, other than the keywords we searched
for to collect this dataset and hashtags we used to label \pro{}
and \anti{} gun control tweets, there was no tailoring of our analysis
to the gun control domain.  This method
of social media analysis can be applied to a wide range of salient
public policy issues.

\bibliographystyle{abbrv}
\bibliography{guncontrol_draft,policy}

\begin{thebibliography}{10}

\bibitem{Ayers:2016uo}
J.~W. Ayers, B.~M. Althouse, E.~C. Leas, T.~Alcorn, and M.~Dredze.
\newblock Big media data can inform gun violence prevention.
\newblock In {\em Bloomberg Data for Good Exchange}, 2016.

\bibitem{barry2013}
C.~L. Barry, E.~E. McGinty, J.~S. Vernick, and D.~W. Webster.
\newblock After newtown -- public opinion on gun policy and mental illness.
\newblock {\em New England journal of medicine}, 368(12):1077--1081, 2013.

\bibitem{Benton:2016dn}
A.~Benton, M.~J. Paul, B.~Hancock, and M.~Dredze.
\newblock Collective supervision of topic models for predicting surveys with
  social media.
\newblock In {\em Association for the Advancement of Artificial Intelligence
  (AAAI)}, 2016.

\bibitem{bermingham2011}
A.~Bermingham and A.~F. Smeaton.
\newblock On using twitter to monitor political sentiment and predict election
  results.
\newblock In {\em IJCNLP Workshop on Sentiment Analysis where AI meets
  Psychology}, 2011.

\bibitem{blei2003}
D.~M. Blei, A.~Y. Ng, and M.~I. Jordan.
\newblock Latent dirichlet allocation.
\newblock {\em J. Mach. Learn. Res.}, 3:993--1022, Mar. 2003.

\bibitem{burstein2003}
P.~Burstein.
\newblock The impact of public opinion on public policy: A review and an
  agenda.
\newblock {\em Political Research Quarterly}, 56(1):29--40, 2003.

\bibitem{Dredze:2016db}
M.~Dredze, D.~A. Broniatowski, M.~Smith, and K.~M. Hilyard.
\newblock Understanding vaccine refusal: Why we need social media now.
\newblock {\em American Journal of Preventive Medicine}, 2015.

\bibitem{dredze2013a}
M.~Dredze, M.~J. Paul, S.~Bergsma, and H.~Tran.
\newblock Carmen: A twitter geolocation system with applications to public
  health.
\newblock In {\em AAAI Workshop on Expanding the Boundaries of Health
  Informatics Using AI (HIAI)}, 2013.

\bibitem{gallup2001}
A.~M. Gallup.
\newblock {\em The Gallup Poll: Public Opinion, 2000}.
\newblock Rowman \& Littlefield, 2001.

\bibitem{ginsberg2009detecting}
J.~Ginsberg, M.~H. Mohebbi, R.~S. Patel, L.~Brammer, M.~S. Smolinski, and
  L.~Brilliant.
\newblock Detecting influenza epidemics using search engine query data.
\newblock {\em Nature}, 457(7232):1012--1014, 2009.

\bibitem{griffiths2004}
T.~L. Griffiths and M.~Steyvers.
\newblock Finding scientific topics.
\newblock {\em Proceedings of the National Academy of Sciences}, 101(suppl
  1):5228--5235, 2004.

\bibitem{grimmer2010}
J.~Grimmer.
\newblock A bayesian hierarchical topic model for political texts: Measuring
  expressed agendas in senate press releases.
\newblock {\em Political Analysis}, 18(1):1--35, 2010.

\bibitem{hong2010}
L.~Hong and B.~D. Davison.
\newblock Empirical study of topic modeling in twitter.
\newblock In {\em KDD Workshop on Social Media Analytics}, pages 80--88, 2010.

\bibitem{jiang2011target}
L.~Jiang, M.~Yu, M.~Zhou, X.~Liu, and T.~Zhao.
\newblock Target-dependent twitter sentiment classification.
\newblock In {\em Association for Computational Linguistics (ACL)}, 2011.

\bibitem{kleck2009}
G.~Kleck, M.~Gertz, and J.~Bratton.
\newblock Why do people support gun control?: Alternative explanations of
  support for handgun bans.
\newblock {\em Journal of Criminal Justice}, 37(5):496--504, 2009.

\bibitem{mccallum2002}
A.~K. McCallum.
\newblock Mallet: A machine learning for language toolkit.
\newblock http://www.cs.umass.edu/~mccallum/mallet, 2002.

\bibitem{mimno2012}
D.~M. Mimno.
\newblock Computational historiography: Data mining in a century of classics
  journals.
\newblock {\em JOCCH}, 5(1):3, 2012.

\bibitem{monroe1998}
A.~D. Monroe.
\newblock Public opinion and public policy, 1980-1993.
\newblock {\em Public Opinion Quarterly}, pages 6--28, 1998.

\bibitem{newman2006}
D.~J. Newman and S.~Block.
\newblock Probabilistic topic decomposition of an eighteenth-century american
  newspaper.
\newblock {\em J. Am. Soc. Inf. Sci. Technol.}, 57(6):753--767, Apr. 2006.

\bibitem{oconnor2010}
B.~O'Connor, R.~Balasubramanyan, B.~R. Routledge, and N.~A. Smith.
\newblock From tweets to polls: Linking text sentiment to public opinion time
  series.
\newblock In {\em International Conference on Weblogs and Social Media
  (ICWSM)}, 2010.

\bibitem{page1983}
B.~I. Page and R.~Y. Shapiro.
\newblock Effects of public opinion on policy.
\newblock {\em The American Political Science Review}, pages 175--190, 1983.

\bibitem{paul2011}
M.~J. Paul and M.~Dredze.
\newblock You are what you tweet: Analyzing twitter for public health.
\newblock In {\em International Conference on Weblogs and Social Media
  (ICWSM)}, 2011.

\bibitem{paul2012}
M.~J. Paul and M.~Dredze.
\newblock Experimenting with drugs (and topic models): Multi-dimensional
  exploration of recreational drug discussions.
\newblock In {\em AAAI 2012 Fall Symposium on Information Retrieval and
  Knowledge Discovery in Biomedical Text}, 2012.

\bibitem{ramage2010}
D.~Ramage, S.~T. Dumais, and D.~J. Liebling.
\newblock Characterizing microblogs with topic models.
\newblock In {\em International Conference on Weblogs and Social Media
  (ICWSM)}, 2010.

\bibitem{sang2012}
E.~T.~K. Sang and J.~Bos.
\newblock Predicting the 2011 dutch senate election results with twitter.
\newblock In {\em EACL Workshop on Semantic Analysis in Social Media}, pages
  53--60, 2012.

\bibitem{schildkraut2014}
J.~Schildkraut and T.~C. Hernandez.
\newblock Laws that bit the bullet: A review of legislative responses to school
  shootings.
\newblock {\em American Journal of Criminal Justice}, 39(2):358--374, 2014.

\bibitem{stefanone2015}
M.~A. Stefanone, G.~D. Saxton, M.~J. Egnoto, W.~Wei, and Y.~Fu.
\newblock Image attributes and diffusion via twitter: The case of\# guncontrol.
\newblock In {\em System Sciences (HICSS), 2015 48th Hawaii International
  Conference on}, pages 1788--1797. IEEE, 2015.

\bibitem{tumasjan2010}
A.~Tumasjan, T.~O. Sprenger, P.~G. Sandner, and I.~M. Welpe.
\newblock Predicting elections with twitter: What 140 characters reveal about
  political sentiment.
\newblock In {\em International Conference on Weblogs and Social Media
  (ICWSM)}, 2010.

\bibitem{vernick1993}
J.~S. Vernick, S.~P. Teret, K.~A. Howard, M.~D. Teret, and G.~J. Wintemute.
\newblock Public opinion polling on gun policy.
\newblock {\em Health Affairs}, 12(4):198--208, 1993.

\bibitem{vijayaraghavan2016deepstance}
P.~Vijayaraghavan, I.~Sysoev, S.~Vosoughi, and D.~Roy.
\newblock Deepstance at semeval-2016 task 6: Detecting stance in tweets using
  character and word-level cnns.
\newblock {\em arXiv preprint arXiv:1606.05694}, 2016.

\bibitem{vizzard2015}
W.~J. Vizzard.
\newblock The current and future state of gun policy in the united states.
\newblock {\em Journal of Criminal Law and Criminology}, 104(4):879 -- 904,
  2015.

\bibitem{zhao2011}
W.~Zhao, J.~Jiang, J.~Weng, J.~He, E.-P. Lim, H.~Yan, and X.~Li.
\newblock Comparing twitter and traditional media using topic models.
\newblock In P.~Clough, C.~Foley, C.~Gurrin, G.~Jones, W.~Kraaij, H.~Lee, and
  V.~Mudoch, editors, {\em Advances in Information Retrieval}, volume 6611 of
  {\em Lecture Notes in Computer Science}, pages 338--349. Springer Berlin
  Heidelberg, 2011.

\end{thebibliography}

\end{document}